\documentclass[conference]{IEEEtran}
\IEEEoverridecommandlockouts
\usepackage{cite}
\usepackage{amsmath,amssymb,amsfonts}
\usepackage{graphicx}
\usepackage{textcomp}
\usepackage{xcolor}
\usepackage{hhline}
\usepackage{algorithm}
\usepackage{algpseudocode}
\usepackage{bm}

\def\BibTeX{{\rm B\kern-.05em{\sc i\kern-.025em b}\kern-.08em
    T\kern-.1667em\lower.7ex\hbox{E}\kern-.125emX}}
\begin{document}

\title{Hawkes Process for Understanding the Influence of Pathogenic Social Media Accounts\\
}

\author{\IEEEauthorblockN{Hamidreza Alvari, Paulo Shakarian}
\textit{Arizona State University}\\
Tempe, USA \\
\{halvari, shak\}@asu.edu}

\maketitle

\begin{abstract}
Over the past years, political events and public opinion on the Web have been allegedly manipulated by accounts dedicated to spreading disinformation and performing malicious activities on social media. These accounts hereafter referred to as ``Pathogenic Social Media  (PSM)" accounts, are often controlled by terrorist supporters, water armies or fake news writers and hence can pose threats to social media and general public. Understanding and analyzing PSMs could help social media firms devise sophisticated and automated techniques that could be deployed to stop them from reaching their audience and consequently reduce their threat. In this paper, we leverage the well-known statistical technique ``Hawkes Process" to quantify the influence of PSM accounts on the dissemination of malicious information on social media platforms. Our findings on a real-world ISIS-related dataset from Twitter indicate that PSMs are significantly different from regular users in making a message viral. Specifically, we observed that PSMs do not usually post URLs from mainstream news sources. Instead, their tweets usually receive large impact on audience, if contained URLs from Facebook and alternative news outlets. In contrary, tweets posted by regular users receive nearly equal impression regardless of the posted URLs and their sources. Our findings can further shed light on understanding and detecting PSM accounts.
\end{abstract}


\section{Introduction}
Online social media play major role in dissemination of information. However, recent years have witnessed evidence of spreading huge amount of harmful disinformation on social media and manipulating public opinion on the Web, attributed to accounts dedicated to spreading malicious information. These accounts are referred to ``Pathogenic Social Media'' (PSM) accounts and can pose threats to social media firms and general public. PSMs are usually controlled by terrorist supporters, water armies or fake news writers and they are owned by either real users or bots who seek to promote or degrade certain ideas by utilizing large online communities of supporters to reach their goals. Identifying PSM accounts could have immediate applications, including countering terrorism~\cite{khader2016combating,KlausenMZ16}, fake news detection~\cite{Gupta2014,6805772} and water armies detection~\cite{DBLP:journals/corr/abs-1111-4297,DBLP:conf/asunam/ChenWSB13}.

To better understand the behavior and impact of PSM accounts on the Web and normal users, and be able to counter their malicious activity, social media authorities need to deploy certain capabilities which could ultimately lead to reducing their threats. To make this happen, social media platforms are required to design sophisticated techniques that could automatically detect and suspend these accounts as quickly as possible, before they can reach their vast audience and spread malicious content. However, for the most part, the social media firms usually rely on reports they receive from their normal users or even their assigned teams to manually shut down these accounts. First of all, this mechanism is not always feasible due to the limited manpower and since not many real users are willing to put aside time and report the malicious activities. Also, it cannot be done in a timely manner since it takes time for these firms to review the reports and decide whether they are legit or not. On the other hand, the fact that these accounts simply return to social media using different accounts or even migrate to other social media, makes all these efforts almost useless. Therefore, the burden falls to automatic approaches that can identify these malicious actors on social media.

\textbf{Present work.} In this work, to address the above mentioned challenges, we aim to understand PSM accounts by (1) analyzing their behavior in terms of their posted URLs, and (2) estimate their influence by conducting experiments on a real-world dataset from Twitter. We deploy a mathematical technique known as ``Hawkes process"~\cite{hawkes71}  to quantify the impact of PSMs on normal users and the greater Web, by looking at their posted URLs on Twitter. Hawkes processes are special forms of point processes and have shown promising results in many problems that require modeling complicated event sequences where historical events have impact on future ones, including financial analysis~\cite{bacry2016estimation}, seismic analysis~\cite{daley2007introduction} and social network modeling~\cite{zhou2013learning} to name a few. This study uses an ISIS-related dataset from Twitter~\cite{alvari2018early}. The dataset contains an \textit{action log} of users in the form of cascades of retweets. In this work, we consider URLs posted by two groups of users: (1) PSM accounts and (2) normal users. The URLs can belong to any platform including the major social media (e.g., facebook.com), mainstream news (e.g., nytimes.com) and alternative news outlets (e.g., rt.com). For each group of users, we fit a multi-dimensional Hawkes processes model wherein each process correspond to a platform referenced in at least one tweet. Furthermore, every process can influence all the others including itself, which allows estimating the strength of connections between each of the social media platforms and news sources, in terms of how likely an event (i.e., the posted URL) can cause subsequent events in each of the groups. In other words, in this study we are interested to investigate if a given URL $u_1$ has influence on another URL $u_2$ (i.e., $u_1\rightarrow u_2$) and thus can trigger subsequent events.


\textbf{Main Findings.} This paper makes the following main observations:
\begin{itemize}
	
	\item Among all platforms studied here, URLs shared from facebook.com and alternative news media contributed the most to the dissemination of malicious information from PSM accounts. Simply put, they had the largest impact on making a message viral and causing the subsequent events. 
	
	\item Posts that were tweeted by the PSM accounts and contained URLs from facebook.com, demonstrated more influence on the subsequent retweets containing URLs from youtube.com, in contrary to the other way around. This means that ultimately tweets with URLs from Facebook will high likely end up inducing more external impulse on YouTube than YouTube might have on Facebook.
	
	\item URLs posted by the normal users have nearly the same impact on the subsequent events regardless of the social media or news outlet used. This basically means that normal users do not often prefer specific social media or news sources over the others.
\end{itemize}


\section{Data Analysis}
In this section, we first explain the dataset and provide the list of the main social media platforms and news sites used in this study. Finally, we present our data analysis to demonstrate the differences between PSM accounts and normal users.

\subsection{Dataset}
We collect a dataset of 2.8M ISIS related tweets/retweets in Arabic 
between February 22, 2016 and May 27, 2016. The dataset contains different fields including user ID, retweet ID, hashtags, content, posting time. The dataset also contains user profile information including name, number of followers/followees, description, location, etc. 
The tweets were collected using different ISIS-related hashtags such as \textsf{\#stateoftheislamiccaliphate}. In this dataset, about 600K tweets have at least one URL (i.e., event) referencing one of the social media platforms or news outlets. There are about 1.4M of paired URLs which we denote by $u_1\rightarrow u_2$ and indicates a retweet (with the URL $u_2$) of the original tweet (with the URL $u_1$). 

\begin{figure}[t]\center
	\includegraphics[width=0.3\textwidth]{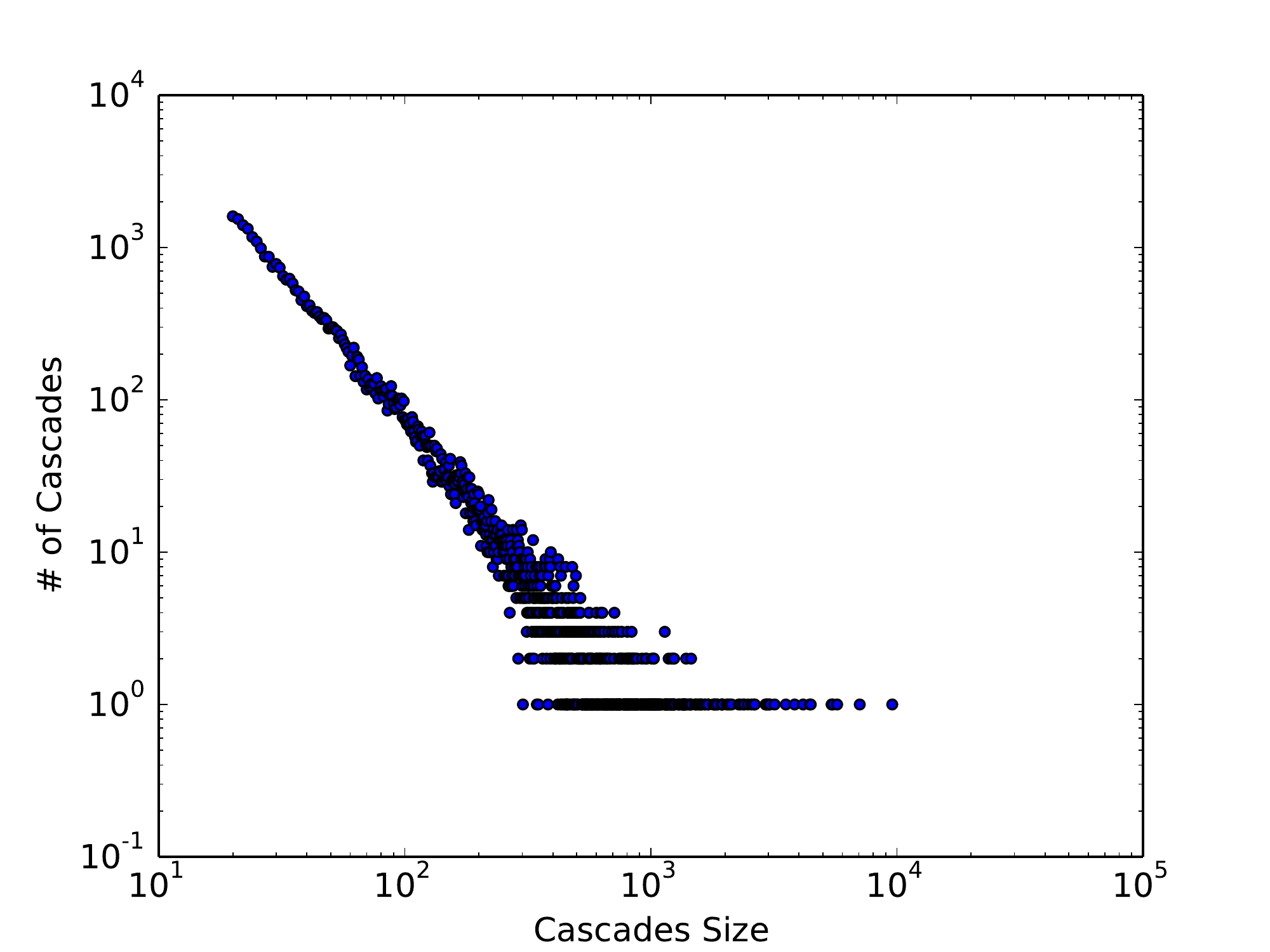}
	\includegraphics[width=0.3\textwidth]{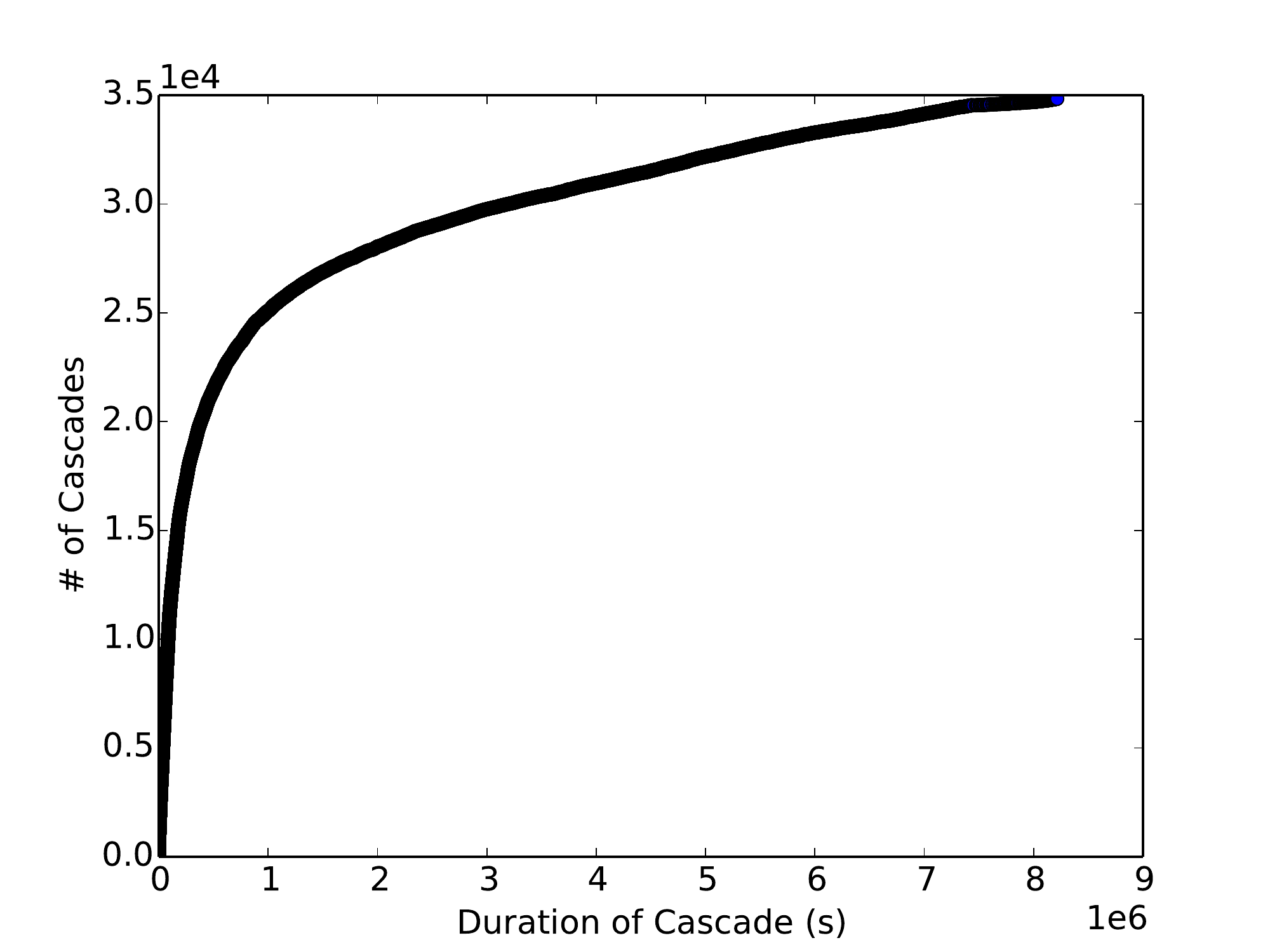}	
	\caption{(Top) Log-log distribution of cascades vs. cascade size. (Bottom) Cumulative distribution of duration of cascades}
	\label{fig:casc_dist}
\end{figure}

In this study, we are interested in investigating the impact of the URL $u_1$ on $u_2$. Accordingly, the dataset contains 35K cascades (i.e., sequences of events) of different sizes and durations, some of which contain paired URLs in the aforementioned form. After pre-processing and removing duplicate users from cascades (those who retweet themselves multiple times), cascades sizes (i.e. number of associated postings) vary between 20 to 9,571 and take from 10 seconds to 95 days to finish. The log-log distribution of cascades vs. cascade size and the cumulative distribution of duration of cascades are depicted in Figure~\ref{fig:casc_dist}. 

\begin{figure}[t]\center
\includegraphics[width=0.3\textwidth]{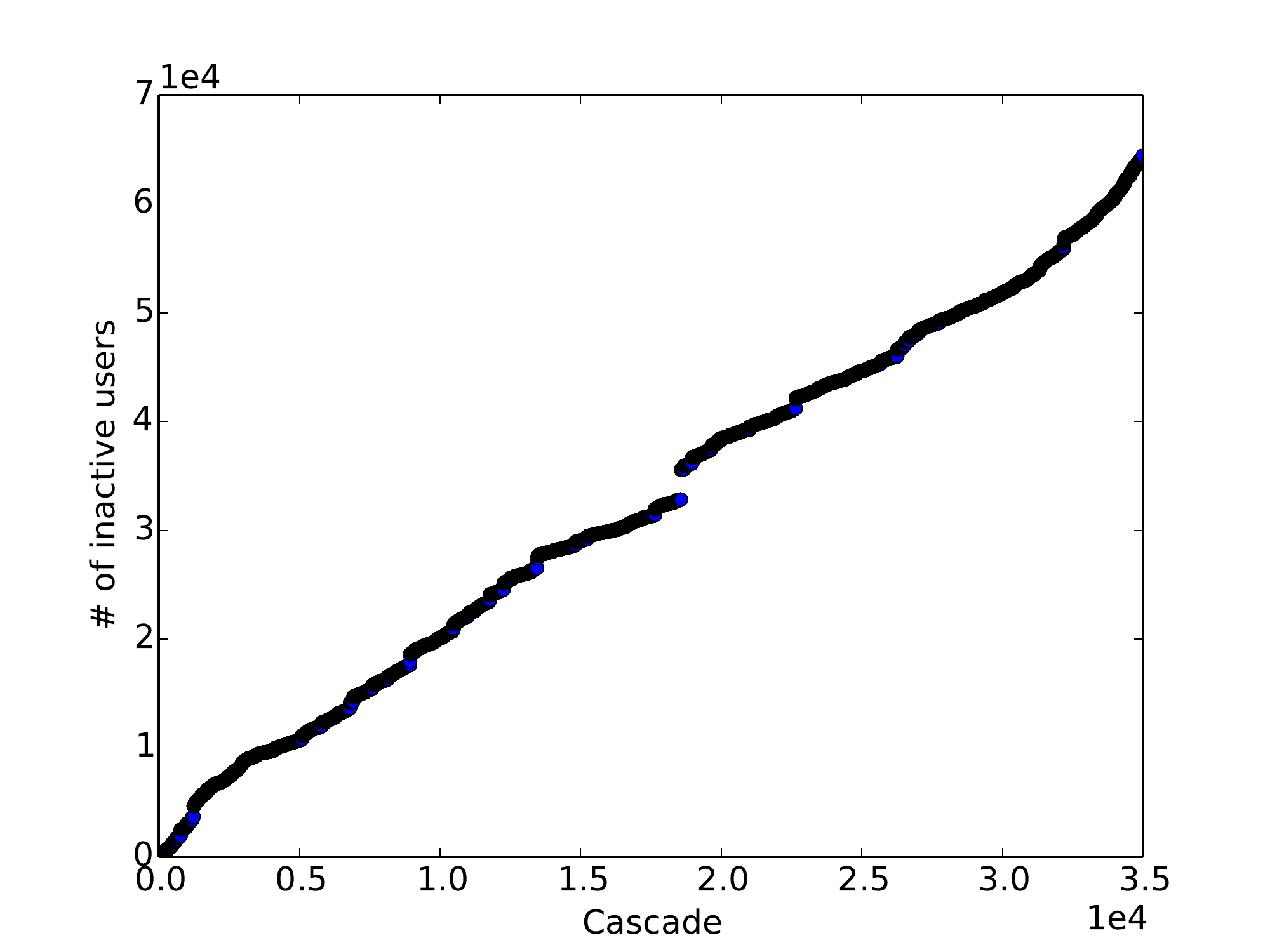}
\caption{Total number of inactive users in each cascade.}
\label{fig:inactive_time}
\end{figure}

The statistics of the dataset are presented in Table~\ref{tb:st}. For labeling, we check through Twitter API to examine whether the users have been suspended (labeled as PSM) or they are still active (labeled as normal)~\cite{thomas2011suspended}. According to Table~\ref{tb:st}, 11\% of the users in our dataset are PSMs and others are normal. We also depict the total number of PSM accounts that have been suspended by Twitter in each cascade, in Figure~\ref{fig:inactive_time}. Finally, to reiterate, we note that these accounts mostly get suspended manually by Twitter based on reports the platform receives from its own users\footnote{https://blog.twitter.com/official/en\_us/a/2016/an-update-on-our-efforts-to-combat-violent-extremism.html}. 

\begin{table}[b]
	\centering
	\caption{Description of the dataset.}
	\begin{tabular}{l|c|c}
		\cline{1-3}
		\textbf{Name}          & \multicolumn{2}{c}{\textbf{Value}}\\
		\cline{1-3}
		\# of Cascades      & \multicolumn{2}{c}{35K}  \\ \cline{1-3}
		\# of Tweets/Retweets & \multicolumn{2}{c}{2.8M} \\ \cline{1-3}
		 & \textbf{PSM} & \textbf{Normal} \\ \cline{2-3}
		\# of Users & 64,484 & 536,609\\
		\cline{1-3}
		\# of Single URLs & 104,948 & 536,046 \\ \cline{1-3}
		\# of Paired URLs & 200,892 & 1,123,434
		\\ \cline{1-3}		
	\end{tabular}
	\label{tb:st}
\end{table}

\subsection{Social Media Platforms and News Outlets}
Twitter deploys a URL shortener technique to leave more space for content and protect users from malicious sites\footnote{https://help.twitter.com/en/using-twitter/url-shortener}. To obtain the original URLs, we use a URL unshortening tool\footnote{https://github.com/skevas/unshorten} to obtain the original links contained in the tweets in our dataset. 

We consider a number of major and well-known social media platforms including Twitter, Facebook, Instagram, Google and Youtube. 
About the dichotomy of mainstream and alternative media, it is notable to mention that most criteria for determining whether a news source counts as either of them, are based on a number of factors including but not limited to the content and whether or not it is corporate owned\footnote{https://smallbusiness.chron.com/mainstream-vs-alternative-media-21113.html}. However, a key difference between these two sources of media comes from the fact that all of mainstream media is profit-oriented, in contrast to the alternative media. We further note that for the most part, mainstream media is considered as a more credible source than alternative media, although the reputation has been recently tainted by the fake news. 

In this work, following the commonsense, we consider popular news outlets such as The New York Times, and The Wall Street Journal as mainstream and less popular ones as alternatives. In Table~\ref{tb:urls_social}, we summarize the total number of paired URLs (i.e, $u_1\rightarrow u_2$) in which the original URL (i.e., $u_1$) corresponds to each social media platform with at least one event in our dataset. We also summarize in Table~\ref{tb:urls_news}, the total number of paired URLs whose original URL belongs to the mainstream and alternative news sources. In Table~\ref{tb:psm_normal_urls}, we see the break down of number of paired URLs for the PSM and normal users. We further demonstrate in Table~\ref{tb:percentage} some examples of the mainstream and alternative news URLs occurrence used in this work.

\begin{table}
	\centering
	\caption{Social media platform's total number of paired URLs of the form $u_1\rightarrow u_2$ with at least one event in the dataset for the PSM and normal users.}
	\begin{tabular}{c|c|c} 
		\cline{1-3}
		\textbf{Platform}        & \textbf{PSM} & \textbf{Normal}\\
		\cline{1-3}
		Twitter & 139,940 & 918,803\\ \hline
		Facebook & 878 & 4,017 \\ \hline
		Instagram & 0 & 2,857\\ \hline
		Google & 163 & 132\\ \hline
		Youtube & 24,724 & 72,890\\ \hline	
	\end{tabular}
	\label{tb:urls_social}
\end{table}

\begin{table}
	\centering
	\caption{News sources' total paired URLs ($u_1\rightarrow u_2$) with at least one event in the dataset for the PSM and normal users.}
	\begin{tabular}{c|c|c} 
		\cline{1-3}
		\textbf{News Source}          & \textbf{PSM} & \textbf{Normal}\\
		\cline{1-3}
		Mainstream & 0 & 286 \\ \hline
		Alternatives & 35,187 &  124,449 \\ \hline		
	\end{tabular}
	\label{tb:urls_news}
\end{table}

\begin{table*}
	\centering
	\caption{Total number of paired URLs of the form $u_1\rightarrow u_2$ with at least one event for PSM/normal users and for all platforms.}
	\begin{tabular}{c|c|c|c|c|c|c|c} 
		\cline{1-8}
		& $\rightarrow$ Twitter  & $\rightarrow$ Facebook & $\rightarrow$ Instagram & $\rightarrow$ Google & $\rightarrow$ Youtube & $\rightarrow$ Mainstream & $\rightarrow$ Alternatives\\
		\cline{1-8}
		Twitter $\rightarrow$ & 109,354/766,617 & 598/3,843 & 229/2,461 & 120/382 & 11,992/59,889 & 90/688 & 17,557/84,923\\ \hline
		Facebook $\rightarrow$& 655/3,108 & 4/41 & 3/9 & 2/1 & 87/281 & 0/1 & 127/576 \\ \hline
		Instagram $\rightarrow$& 0/2,362 & 0/11 & 0/25 & 0/2 & 0/161 & 0/2 & 0/294\\ \hline
		Google $\rightarrow$& 134/74 & 0/0 & 0/1 & 0/0 & 12/53 & 0/0 & 17/4\\ \hline
		Youtube $\rightarrow$& 14,004/56,545 & 132/312 & 23/211 & 22/32 & 6,799/7,529 & 13/48 & 3,731/8,213\\ \hline
		Mainstream $\rightarrow$& 0/189 & 0/1 & 0/1 & 0/0 & 0/13 & 0/1 & 0/81\\ \hline	
		Alternatives $\rightarrow$& 21,047/95,641 & 145/767 & 45/318 & 59/64 & 3,862/9,199 & 26/122 & 10,003/18,338\\ \hline		
	\end{tabular}
	\label{tb:psm_normal_urls}
\end{table*}

\begin{table}[b]
	\centering
	\caption{Examples of mainstream and alternative news. 
		}
	\begin{tabular}{c|c}
		\cline{1-2}
		\textbf{Mainstream} & \textbf{Alternatives} \\ 
		\cline{1-2}
		    https://www.nytimes.com  & https://www.rt.com  \\ 
		    https://www.reuters.com  &  https://www.arabi21.com \\ 
		    https://www.wsj.com  &  https://www.7adramout.net \\ 
		    https://www.nbcnews.com  &  https://www.addiyar.com \\
		    https://www.ft.com & https://zamnpress.com \\
		\cline{1-2}
	\end{tabular}
	\label{tb:percentage}
\end{table}

\subsection{Temporal Analysis}
Here, we present the differences between the PSM accounts in our dataset with their counterparts, normal users through temporal analysis of their posted URLs. 

\begin{figure}[t]\center
	\includegraphics[width=0.4\textwidth]{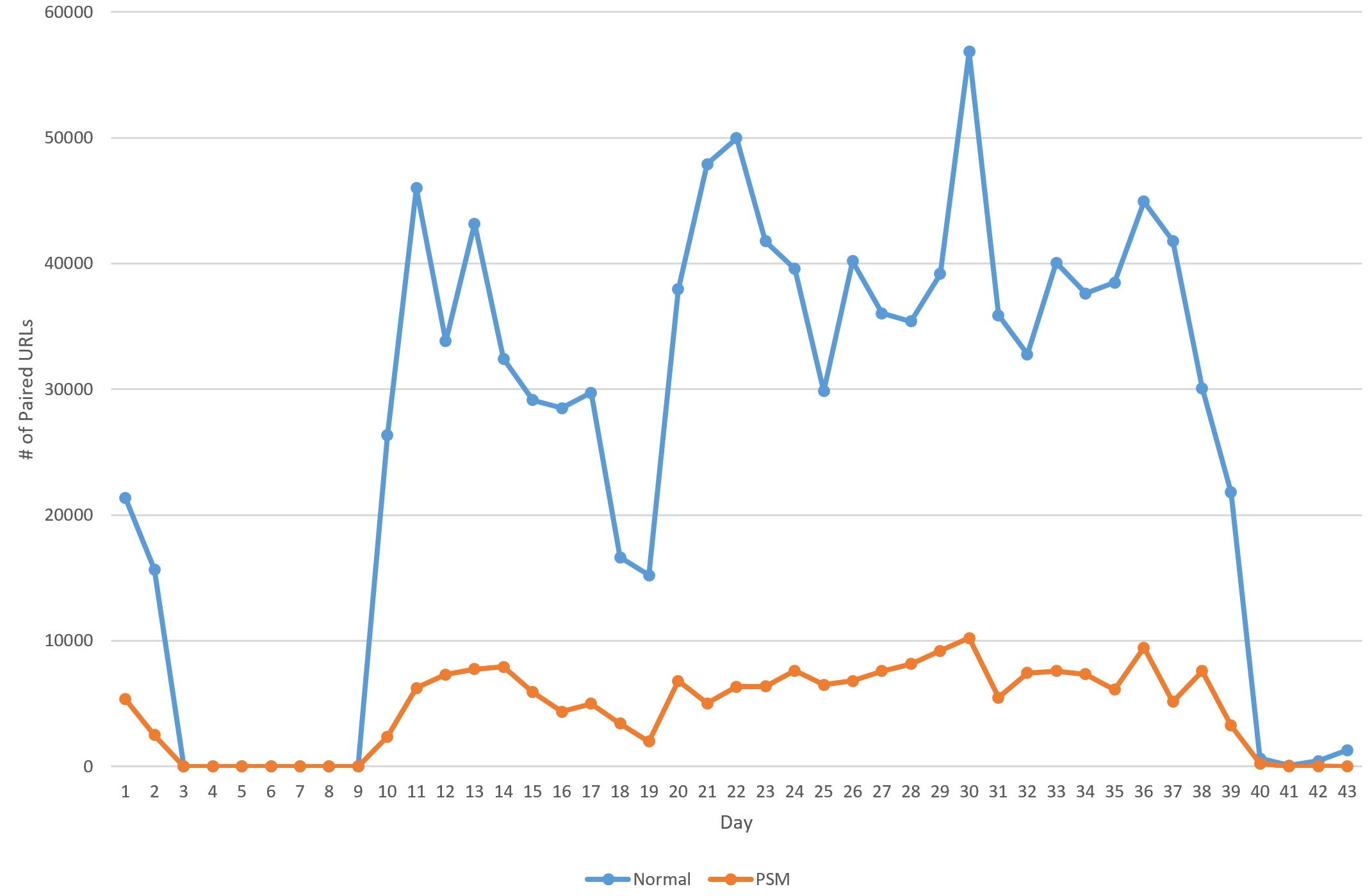}
	\caption{Number of paired URLs posted by the PSM and normal users in our dataset. Note that number of normal users in our dataset is higher than the PSM accounts.}
	\label{fig:temporal}
\end{figure}

In Figure~\ref{fig:temporal}, we depict the daily occurrence of the paired URLs over the span of 43 days for both PSM and normal users. Recall from the previous section that our dataset has a larger number of normal users and higher number of the posted URLs compared to the PSM accounts. Therefore, it is reasonable to observe more activity from normal users than PSMs. For both groups of users, we observe a similar trend in occurrence of spikes and their durations. 
As it is seen, distinguishing between PSMs and normal users merely based on the occurrence of URLs and their patterns is not reliable. Therefore, we set out to conduct experiments using a more sophisticated statistical technique known as ``Hawkes Process" in the next section.

\section{Framework}
In the previous section, we presented our data analysis to demonstrate differences between PSM accounts and normal users in terms of URLs they usually post on Twitter. We now set out to assess their impact via a well-known mathematical technique called ``Hawkes process". 
 
\subsection{Hawkes Processes}
In many scenarios, one needs to deal with timestamped events such as the activity of users on a social network recorded in continuous time. An important task then is to estimate the influence of the nodes based on their timestamp patterns~\cite{gomez2013modeling}. Point process is a principled framework for modeling such event data, where the dynamic of the point process can be captured by its conditional intensity function as follows:

\begin{equation}
\lambda(t) = \lim\limits_{\Delta t \rightarrow 0} \frac{\mathbb{E}(N(t+\Delta t) - N(t)|\mathcal{H}_t)}{\Delta t} = \frac{\mathbb{E}(dN(t)|\mathcal{H}_t)}{dt}
\end{equation}   

\noindent where $\mathbb{E}(dN(t)|\mathcal{H}_t)$ is the expectation of the number of events happend in the interval $(t,t+dt]$ given the historical observations $\mathcal{H}_t$ and $N(t)$ records the number of events before time $t$. Point process can be equivalently represented as a counting process $N=\{N(t)|t \in [0,T]\}$ over the time interval $[0, T]$.

The Hawkes process framework~\cite{hawkes71} has been used in many problems that require modeling complicated event sequences where historical events have impact on future ones. Examples include but are not limited to financial analysis~\cite{bacry2016estimation}, seismic analysis~\cite{daley2007introduction} and social network modeling~\cite{zhou2013learning}. One-dimensional Hawkes process is a point process $N_t$ with the following particular form of intensity function:

\begin{equation}
\lambda(t) = \mu + a\int_{-\infty}^{t}g(t-s)dN_s = \mu + a\sum_{i:t_i<t}{g(t-t_i)}
\end{equation}

\noindent where $\mu > 0$ is the exogenous base intensity (i.e., background rate of events), $t_i$ are the time of events in the point process before time $t$, and $g(t)$ is the decay kernel. 

In this paper, we use exponential kernel of the form $g(t) = we^{-wt}$, but adapting to the other positive forms is straightforward. The second part of the above formulation captures the self-exciting nature of the point processes-- the occurrence of events in the past has a positive impact on the future ones. Given a sequence of events $\{t_i\}_{i=1}^n$ observed in $[0, T]$ and generated from the above intensity function, the log-likelihood function can be obtained as follows~\cite{zhou2013learning}:

\begin{equation}\label{eq:log_like}
\mathcal{L} = \log\frac{\prod_{i=1}^n\lambda(t_i)}{\exp\int_{0}^{T}\lambda(t)dt} = \sum_{i=1}^n{\log\lambda(t_i)}-\int_{0}^{T}\lambda(t)dt
\end{equation}

In this paper, we focus on multi-dimensional Hawkes processes which is defined by a $U$-dimensional point process $N_t^u, u=1,\ldots,U$. In other words, we have $U$ Hawkes processes coupled with each other-- each Hawkes process correspond to one of the platforms and the influence between them is modeled using the mutually-exciting property of the multi-dimensional Hawkes processes. We formally define the following formulation to model the influence of different events on each other:

\begin{equation}\label{eq:multi_hawkes}
\lambda_u(t) = \mu_u + \sum_{i:t_i<t}{a_{uu_i}g(t-t_i)}
\end{equation}

\noindent where $\mu_u \geq 0$ is the base intensity for the $u$-th Hawkes process. The coefficient $a_{uu_i}\geq 0$ captures the mutually-exciting property between the $u$-th and $u_i$-th processes. Larger value of $a_{uu_i}$ shows that events in the $u_i$-th dimension are more likely to trigger an event in $u$-th dimension in future. More intuitively, an event on one point process can cause an impulse response on other processes, which increases the probability of an event occurring above the processes' background rates. We reiterate that in this study each URL is attributed to an event, i.e., if the URL $u_1$ triggers the URL $u_2$ (i.e., $u_1\rightarrow u_2$), then $a_{u_2u_1}\geq 0$ 

\begin{figure}[t]\center
	\includegraphics[width=0.4\textwidth]{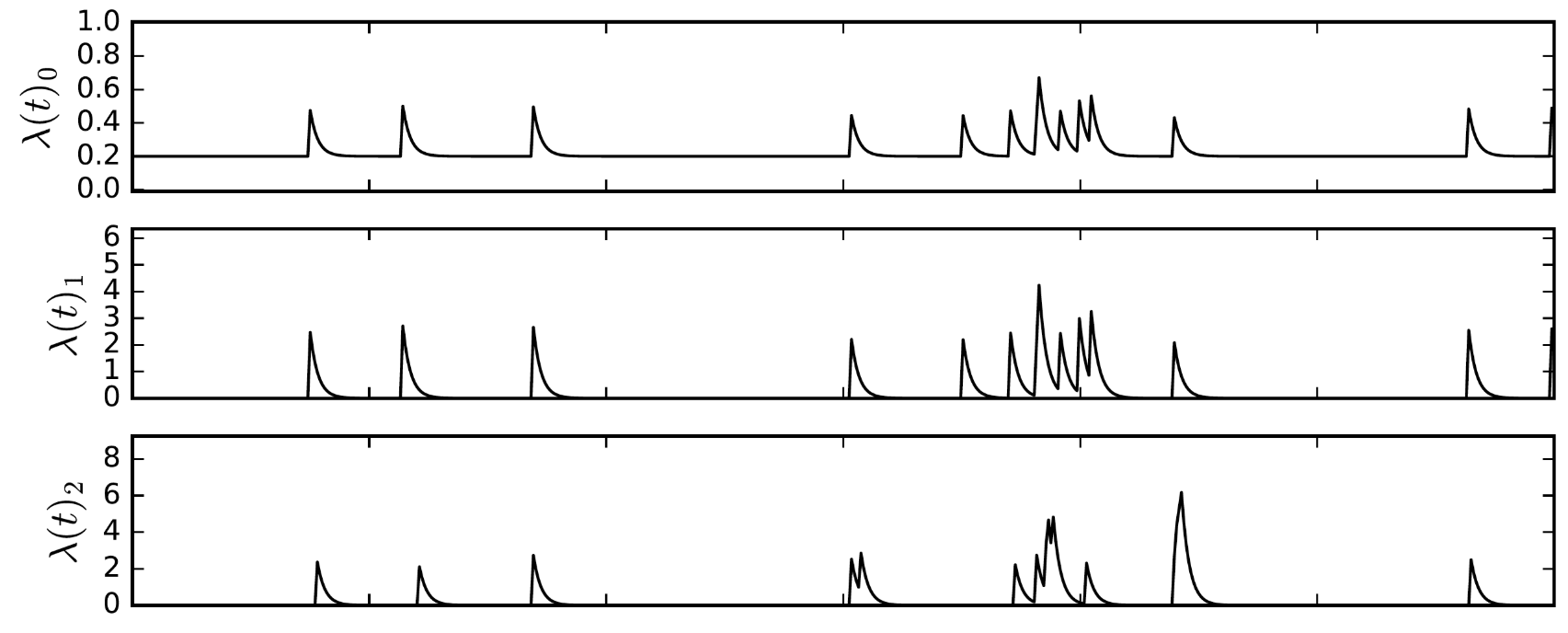}
	\caption{Illustration of the Hawkes Process. Events induce impulse on other processes and cause child events. Background event in $e_0$ induces impulse on responses on processes $e_1$ and $e_2$.}
	\label{fig:hawkes}
\end{figure}

In Figure~\ref{fig:hawkes}, we depict a multivariate example of three different streams of events, $e_0$, $e_1$ and $e_2$. As illustrated, $e_0$ is caused by the background rate $\lambda(t)_0$ and has an influence on itself and $e_1$. On the other hand, $e_1$ is caused by $\lambda(t)_1$ and has an influence on $e_2$. Simply put, a background event in $e_0$ induces impulse on responses on processes $e_1$ and $e_2$. Accordingly, the caused child event in $e_1$ leads to another child event in $e_2$. 

We consider an infectivity matrix $\boldsymbol{A}=[a_{uu_i}] \in \mathbb{R}^{U\times U}$ which collects the self-triggering coefficients between Hawkes processes, and $U=7$ is the number of processes (i.e., platforms) in our work. Each entry in this matrix indicates the strength of influence each platform has on other platforms. Our ultimate goal in this paper is to estimate the infectivity matrix as it reflects the estimated influence of each platform on others. Next, we will provide the methodology that we follow to estimate the influence of the URLs on each other.

\subsection{Methodology}
We aim to assess the influence of the PSM accounts in our dataset via their posted URLs. We consider the URLs posted by two groups of users: (1) PSM accounts and (2) normal users. For both groups, we fit a Hawkes model with $K=7$ point processes each for the seven categories of social media and news outlets discussed earlier. In each of the Hawkes models, every process is able to influence all the others including itself, which allows us to estimate the strength of connections between each of the seven categories for both groups of users, in terms of how likely an event (i.e., the posted URL) can cause subsequent events in each of the groups. 

We use the \textsc{Adm4} algorithm presented by~\cite{zhou2013learning} and follow the methodology presented by~\cite{zannettou2017web} for fitting the Haweks processes for both PSM and normal users. \textsc{Adm4}~\cite{zhou2013learning} is an efficient optimization that estimates the parameters $\boldsymbol{A}$ and $\boldsymbol{\mu}$ by maximizing the regularized log-likelihood $\mathcal{L}(\boldsymbol{A},\boldsymbol{\mu})$:

\begin{equation}
\min_{\boldsymbol{A}\geq 0,\boldsymbol{\mu}\geq 0} -\mathcal{L}(\boldsymbol{A},\boldsymbol{\mu}) + \lambda_1{||\boldsymbol{A}||_*} + \lambda_2||\boldsymbol{A}||_1
\end{equation} 

\noindent where $\mathcal{L}(\boldsymbol{A},\boldsymbol{\mu})$ can be obtained by substituting $\lambda_u(t)$ from Equation~\ref{eq:multi_hawkes} into Equation~\ref{eq:log_like}. Also, $||\boldsymbol{A}||_*$ is the nuclear norm of matrix $\boldsymbol{A}$, and is defined as the sum of its singular value.

We consider two different sets of URLs posted by the PSM accounts and normal users by selecting URLs that have at least one event in Twitter (i.e., posted by a user). For each group, we construct a matrix $\boldsymbol{W}\in\mathbb{N}^{T\times U}$ with $U=7$, whose entries are sequences of events (i.e., posted URLs) observed during a time period $T$. We note that each sequence of events is of the form $\mathcal{S}=\{(t_i,u_i)\}_{i=1}^{n_i}$ where $n_i$ is the number of the events occurring at the $u_i$-th dimension (i.e., URLs posted containing one of the 7 platforms).  


\section{Experimental Results}
Here, we conduct experiments to gauge the effectiveness of Hawkes process for moderling influence of PSMs.
\subsection{Settings}
In this work, we adopt the \textsc{Adm4} algorithm~\cite{zhou2013learning} which implements parametric inference for Hawkes processes with an exponential kernel and a mix of Lasso and nuclear regularization. We initialize infectivity matrix $\boldsymbol{A}$, base intensities $\boldsymbol{\mu}$ and decays $\boldsymbol{\beta}\in\mathbb{R}$ randomly. 

We further set the number of nodes $U=7$ to reflect the $7$ platforms used in this study. Level of penalization is set to $C=1000$, and the ratio of Lasso-Nuclear regularization mixing parameter is set to $0.5$. Finally, maximum number of iterations for solving the optimization is set to $50$ and the tolerance of solving algorithm is set to $1e-5$. 

\subsection{Results}
\begin{figure}[t]\center
	\includegraphics[width=0.4\textwidth]{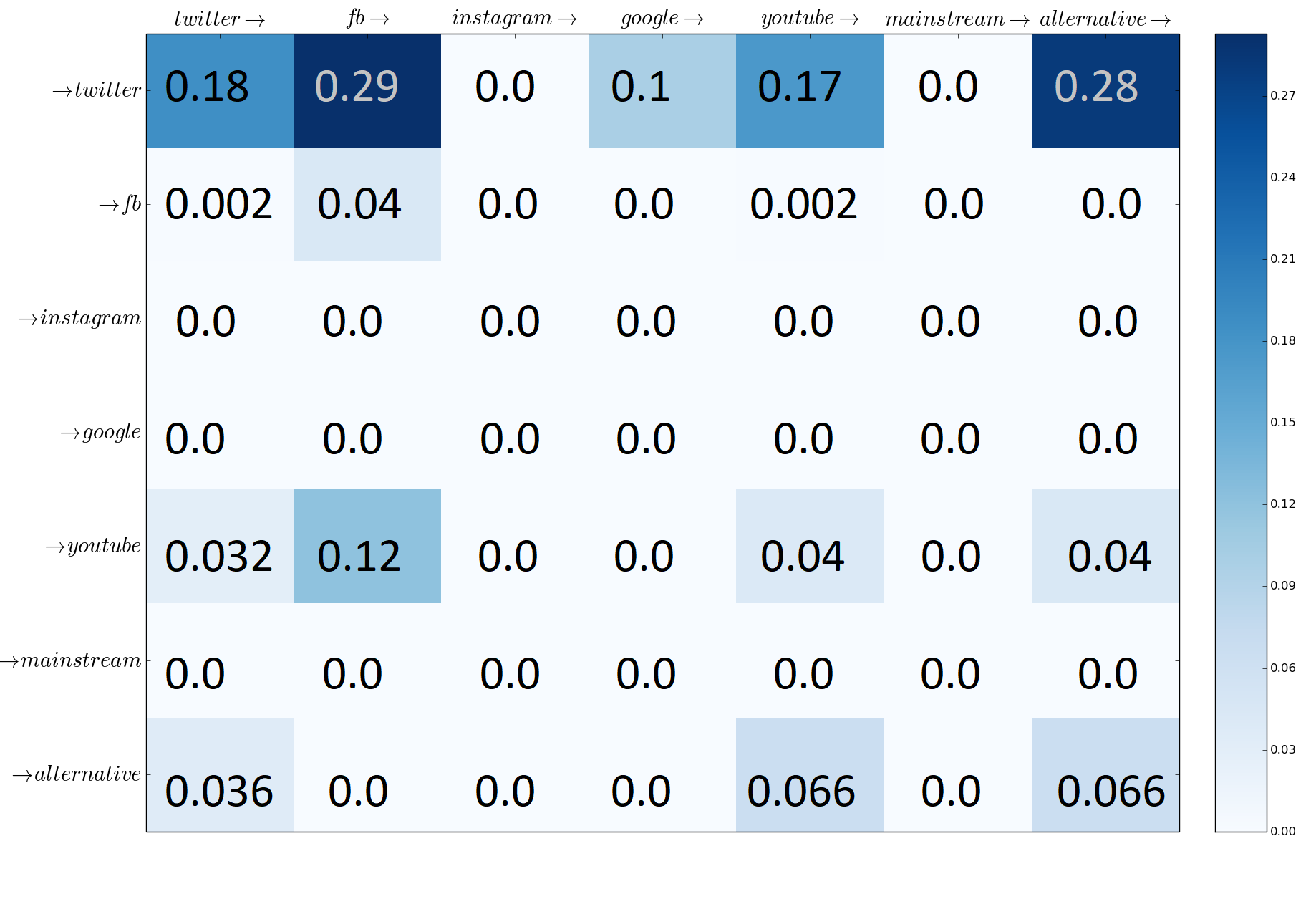}
	\includegraphics[width=0.4\textwidth]{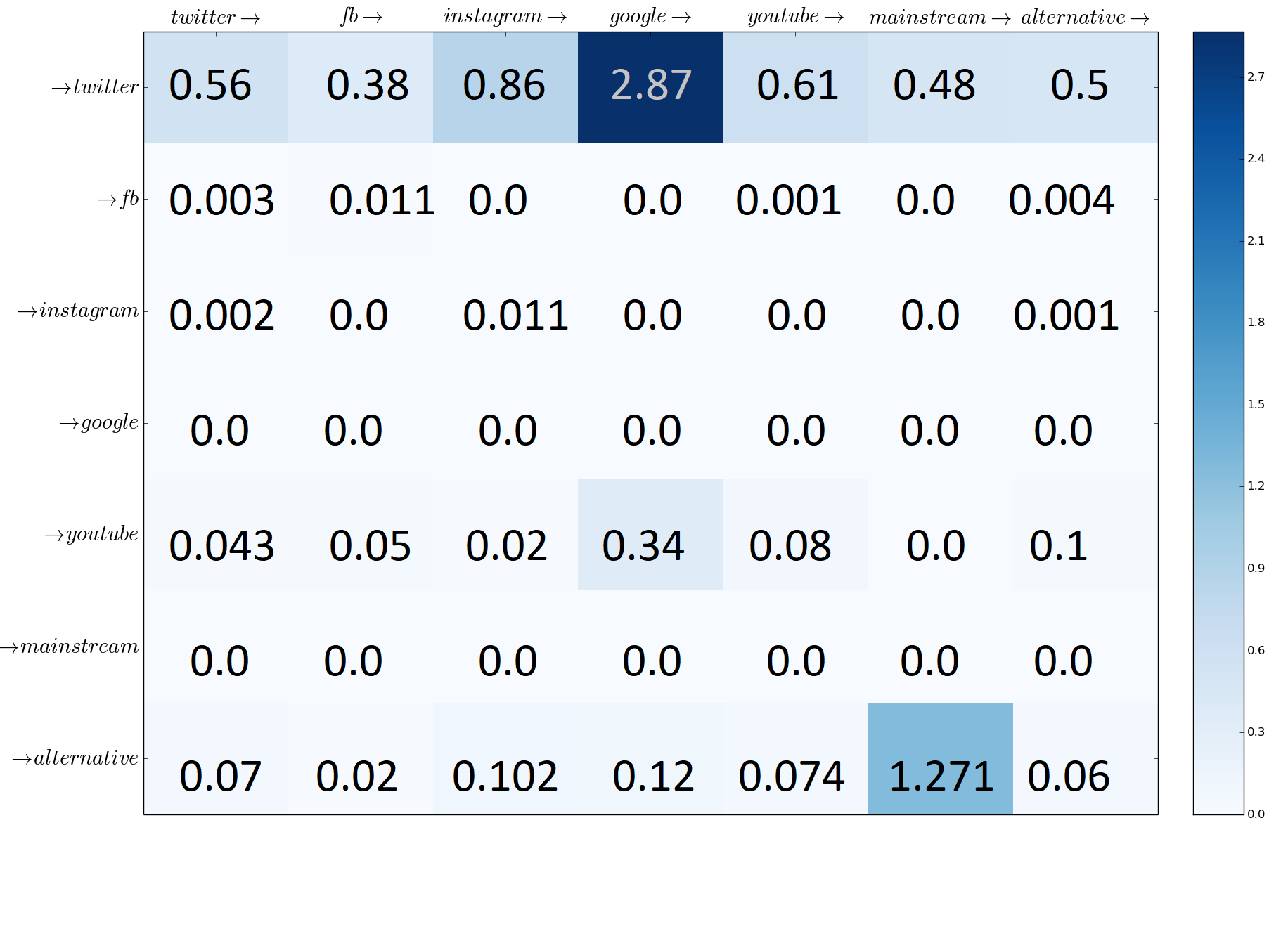}
	\caption{Estimated weights for all paired URLs for (top) PSMs, and (bottom) normal users. Among all URLs, those from facebook.com and alternative news media had the largest impact on dissemination of malicious messages.}
	\label{fig:hawkes_w}
\end{figure}

We estimate infectivity matrix for both PSM and normal users by fitting the Hawkes model described earlier. In our study, this matrix characterizes the strength of the connections between the platforms and news sources. More specifically, each weight value represents the connection strength from one platform to another. In other words, each entry in this matrix can be interpreted as the expected number of subsequent events that will occur on the second group after each event on the first~\cite{zannettou2017web}. In Figure~\ref{fig:hawkes_w}, we depict the estimated weights for all paired URLs for both PSM and normal users. Looking at the weights in both of the plots, we realize that greater weights belong to processes that have impact on Twitter, i.e. ``$\rightarrow Twitter$". This implies that both of the groups in our Twitter dataset often post URLs that ultimately have greater impact on Twitter. 

Overall, we observe the followings:

\begin{itemize}
	\item URLs referencing all platforms and posted by the PSMs and regular users, mostly trigger URLs that contain the Twitter domain.
	\item Among all platforms studied here, URLs shared from facebook.com and alternative news media contributed the most to the dissemination of malicious information from PSM accounts. In other words, they had largest impact on making a message viral and causing the subsequent events. 
	\item Posts that were tweeted by the PSM accounts and contained URLs from facebook.com, demonstrated more influence on the subsequent retweets containing URLs from youtube.com, in contrary to the other way around. This means that ultimately tweets with URLs from facebook will likely end up inducing external impulse on youtube.com. In contrast, URLs posted by the normal users have nearly the same impact on the subsequent events regardless of the social media or news outlet used.
\end{itemize}


The above mentioned observations demonstrate the effectiveness of leveraging Hawkes process to quantify the impact of URLs posted by PSMs and regular users on the dissemination of content on Twitter. The observations we make here show that PSM accounts and regular users behave differently in terms of the URLs they post on Twitter, in that they have different tastes while disseminating URL links. Accordingly their impact on the subsequent events significantly differ from each other.  

\section{Related Work}
The explosive growth of the Web has raised numerous challenges and attracted several researchers with different background~\cite{ferrara2016rise,beigi2019protecting,alvari2016non,beigi2016exploiting,howard2016bots,beigi2014leveraging,alvari2018early,beigi2018privacy,benigni2017online,beigi2018securing,alvari2017semi,wang2014detection,beigi2018similar,alvari2016identifying,ferrara2016predicting,alvari2011detecting}. Below, we will review some of the closest research directions to our research.

\noindent \textbf{Point Process.} When dealing with timestamped events in continuous time such as the activity of users on social media, point process could be leveraged for modeling such events. Point processes have been extensively used to model activities in networks~\cite{xiao2017wasserstein}. Hawkes process is a special form of point processes which models complicated event sequences with historical events influencing future ones. Hawkes processes have been applied to a variety of problems including financial analysis~\cite{bacry2016estimation}, seismic analysis~\cite{daley2007introduction} and social network modeling~\cite{zhou2013learning}, community detection~\cite{tran2015netcodec}, and causal inference~\cite{xu2016learning}.  

\noindent \textbf{Social Spam/Bot Detection.} Social bot is a computer program that automatically generate content and interacts with real people on social media, trying to emulate and possibly alter their behavior~\cite{ferrara2016rise}. Recently, DARPA organized a Twitter bot challenge to detect ``influence bots''~\cite{7490315}, where supervised and semi-supervised approaches were proposed using different features. The work of~\cite{Cao:2014:ULG:2660267.2660269} for example, use similarity to cluster accounts and uncover groups of malicious users. The work of~\cite{varol2017online} presents a supervised framework for bot detection which uses more than thousands features. In a different attempt, the work of~\cite{ICWSM1715678} studied the problem of spam detection in Wikipedia using different spammers behavioral features. 
For a comprehensive survey on the ongoing efforts to fight social bots, we direct the reader to~\cite{ferrara2016rise}.

\noindent \textbf{Fake News Identification.} Fake news detection has recently attracted a growing interest of general public and researchers, as the spread of misinformation on social media and the Web increases on a daily basis. A growing body of work has been devoted to addressing the impact of bots in manipulating political discussion and spreading fake news, including the 2016 U.S. presidential election~\cite{howard2016bots,bessi2016,shao2017spread} and the 2017 French election~\cite{ferrara2017}. For example,~\cite{shao2017spread} analyzes tweets following recent U.S. presidential election and found evidences that bots played key roles in spreading fake news.  

\noindent \textbf{Identifying Instigators.} Given a snapshot of the diffusion process at a given time, these works aim to detect the source of the diffusion. For instance, ~\cite{Zhu:2016:ISD:2942477.2942508} designed an approach for information source detection and in particular initiator of a cascade. In contrast, we are focused on a set of users who \textit{might} or \textit{might not} be initiators. Other similar works on finding most influential spreaders of information such as ~\cite{journals/corr/PeiMAZM14,Fu2015} and outbreak prediction such as ~\cite{Cui:2013:COP:2487575.2487639} also exist in the literature. For example, the work of ~\cite{Konishi:2016:IKO:3061053.3061145} performed classification to detect users who adopt popular items.  

\noindent \textbf{Extremism Detection.} Several studies have focused on understanding extremism in social networks~\cite{benigni2017online,benigni2016tweets,KlausenMZ16,ferrara2016predicting,scanlonautomatic,scanlonforecasting,hung2016detecting}. The work of~\cite{KlausenMZ16}, uses Twitter and proposes an approach to predict new extremists, determine if the newly created account belongs to a suspended extremist, and predict the ego-network of the suspended extremist upon creating her new account.
Authors in~\cite{benigni2016tweets,benigni2017online} performed iterative vertex clustering and classification to identify Islamic Jihadists on Twitter. 

\noindent \textbf{Detection of Internet Water Armies.} The term ``Internet water armies" refers to a special group of online users who get paid for posting comments for some hidden purposes such as influencing other users towards social events or business markets. Therefore, they are also called ``hidden paid posters". The works of~\cite{DBLP:journals/corr/abs-1111-4297,DBLP:conf/asunam/ChenWSB13,wang2014detection} use user behavioral and domain-specific attributes and designed approaches to detect Internet water armies.  


\section{Conclusion and Future Work}
In this study, we presented an analysis on a real-world ISIS-related dataset from Twitter to demonstrate how Pathogenic Social Media (PSM) and normal users usually post on Twitter in terms of the URLs they post. More specifically, we leveraged a statistical technique known as Hawkes Process for modeling the influence of PSM accounts on dissemination of malicious content on the Web. In this work, we used URLs posted by two groups of users, PSMs and normal users, on major social media and mainstream and alternative news outlets. Overall, our findings indicate that the URLs posted by the PSM accounts have the largest impact if contained either facebook.com or alternative news media. In contrast, their counterparts, i.e., normal users, often post URLs that have nearly the same impact on the Web, no matter what social media or news outlet they use. There are potential avenues for future work. First, we would like to extend the study by proposing a prediction mechanism for distinguishing PSMs from normal users, based on their different impact on the greater Web. Another research direction would also be learning causal inference for Hawkes process and investigating the relation between the two concepts, while considering the problem of identification of PSM accounts.  


\section*{Acknowledgment}
This work was supported through DoD Minerva program.


\begin{thebibliography}{10}

\bibitem{khader2016combating}
M.~Khader, \emph{Combating Violent Extremism and Radicalization in the Digital
	Era}, ser. Advances in Religious and Cultural Studies.\hskip 1em plus 0.5em
minus 0.4em\relax IGI Global, 2016.

\bibitem{KlausenMZ16}
J.~Klausen, C.~Marks, and T.~Zaman, ``Finding online extremists in social
networks,'' \emph{CoRR}, vol. abs/1610.06242, 2016.

\bibitem{Gupta2014}
A.~Gupta, P.~Kumaraguru, C.~Castillo, and P.~Meier, \emph{TweetCred: Real-Time
	Credibility Assessment of Content on Twitter}.\hskip 1em plus 0.5em minus
0.4em\relax Springer International Publishing, 2014.

\bibitem{6805772}
A.~Gupta, H.~Lamba, and P.~Kumaraguru, ``\$1.00 per rt \#bostonmarathon
\#prayforboston: Analyzing fake content on twitter,'' in \emph{2013 APWG
	eCrime Researchers Summit}, 2013.

\bibitem{DBLP:journals/corr/abs-1111-4297}
C.~Chen, K.~Wu, S.~Venkatesh, and X.~Zhang, ``Battling the internet water army:
Detection of hidden paid posters,'' \emph{CoRR}, vol. abs/1111.4297, 2011.

\bibitem{DBLP:conf/asunam/ChenWSB13}
C.~Chen, K.~Wu, S.~Venkatesh, and R.~K. Bharadwaj, ``The best answers? think
twice: online detection of commercial campaigns in the {CQA} forums,'' in
\emph{ASONAM}, 2013.

\bibitem{hawkes71}
A.~G. Hawkes, ``Spectra of some self-exciting and mutually exciting point
processes,'' \emph{Biometrika}, vol.~58, no.~1, pp. 83--90, 1971.

\bibitem{bacry2016estimation}
E.~Bacry, T.~Jaisson, and J.-F. Muzy, ``Estimation of slowly decreasing hawkes
kernels: application to high-frequency order book dynamics,''
\emph{Quantitative Finance}, vol.~16, no.~8, pp. 1179--1201, 2016.

\bibitem{daley2007introduction}
D.~J. Daley and D.~Vere-Jones, \emph{An introduction to the theory of point
	processes: volume II: general theory and structure}.\hskip 1em plus 0.5em
minus 0.4em\relax Springer Science \& Business Media, 2007.

\bibitem{zhou2013learning}
K.~Zhou, H.~Zha, and L.~Song, ``Learning social infectivity in sparse low-rank
networks using multi-dimensional hawkes processes,'' in \emph{Artificial
	Intelligence and Statistics}, 2013, pp. 641--649.

\bibitem{alvari2018early}
H.~Alvari, E.~Shaabani, and P.~Shakarian, ``Early identification of pathogenic
social media accounts,'' in \emph{2018 IEEE International Conference on
	Intelligence and Security Informatics (ISI)}.\hskip 1em plus 0.5em minus
0.4em\relax IEEE, 2018, pp. 169--174.

\bibitem{thomas2011suspended}
K.~Thomas, C.~Grier, D.~Song, and V.~Paxson, ``Suspended accounts in
retrospect: an analysis of twitter spam,'' in \emph{Proceedings of the 2011
	ACM SIGCOMM conference on Internet measurement conference}.\hskip 1em plus
0.5em minus 0.4em\relax ACM, 2011, pp. 243--258.

\bibitem{gomez2013modeling}
M.~Gomez-Rodriguez, J.~Leskovec, and B.~Sch{\"o}lkopf, ``Modeling information
propagation with survival theory,'' in \emph{International Conference on
	Machine Learning}, 2013, pp. 666--674.

\bibitem{zannettou2017web}
S.~Zannettou, T.~Caulfield, E.~De~Cristofaro, N.~Kourtelris, I.~Leontiadis,
M.~Sirivianos, G.~Stringhini, and J.~Blackburn, ``The web centipede:
understanding how web communities influence each other through the lens of
mainstream and alternative news sources,'' in \emph{Proceedings of the 2017
	Internet Measurement Conference}.\hskip 1em plus 0.5em minus 0.4em\relax ACM,
2017, pp. 405--417.

\bibitem{ferrara2016rise}
E.~Ferrara, O.~Varol, C.~Davis, F.~Menczer, and A.~Flammini, ``The rise of
social bots,'' \emph{Communications of the ACM}, vol.~59, no.~7, pp. 96--104,
2016.

\bibitem{beigi2019protecting}
G.~Beigi, R.~Guo, A.~Nou, Y.~Zhang, and H.~Liu, ``Protecting user privacy: An
approach for untraceable web browsing history and unambiguous user
profiles,'' in \emph{Proceedings of the Twelfth ACM International Conference
	on Web Search and Data Mining}.\hskip 1em plus 0.5em minus 0.4em\relax ACM,
2019, pp. 213--221.

\bibitem{alvari2016non}
H.~Alvari, P.~Shakarian, and J.~K. Snyder, ``A non-parametric learning approach
to identify online human trafficking,'' in \emph{Intelligence and Security
	Informatics (ISI), 2016 IEEE Conference on}.\hskip 1em plus 0.5em minus
0.4em\relax IEEE, 2016, pp. 133--138.

\bibitem{beigi2016exploiting}
G.~Beigi, J.~Tang, S.~Wang, and H.~Liu, ``Exploiting emotional information for
trust/distrust prediction,'' in \emph{Proceedings of the 2016 SIAM
	international conference on data mining}.\hskip 1em plus 0.5em minus
0.4em\relax SIAM, 2016, pp. 81--89.

\bibitem{howard2016bots}
P.~N. Howard, ``Bots and automation over twitter during the us election,''
2016.

\bibitem{beigi2014leveraging}
G.~Beigi, M.~Jalili, H.~Alvari, and G.~Sukthankar, ``Leveraging community
detection for accurate trust prediction,'' 2014.

\bibitem{beigi2018privacy}
G.~Beigi and H.~Liu, ``Privacy in social media: Identification, mitigation and
applications,'' \emph{arXiv preprint arXiv:1808.02191}, 2018.

\bibitem{benigni2017online}
M.~C. Benigni, K.~Joseph, and K.~M. Carley, ``Online extremism and the
communities that sustain it: Detecting the isis supporting community on
twitter,'' \emph{PloS one}, 2017.

\bibitem{beigi2018securing}
G.~Beigi, K.~Shu, Y.~Zhang, and H.~Liu, ``Securing social media user data: An
adversarial approach,'' in \emph{Proceedings of the 29th on Hypertext and
	Social Media}.\hskip 1em plus 0.5em minus 0.4em\relax ACM, 2018, pp.
165--173.

\bibitem{alvari2017semi}
H.~Alvari, P.~Shakarian, and J.~K. Snyder, ``Semi-supervised learning for
detecting human trafficking,'' \emph{Security Informatics}, vol.~6, no.~1,
p.~1, 2017.

\bibitem{wang2014detection}
K.~Wang, Y.~Xiao, and Z.~Xiao, ``Detection of internet water army in social
network,'' 2014.

\bibitem{beigi2018similar}
G.~Beigi and H.~Liu, ``Similar but different: Exploiting users' congruity for
recommendation systems,'' in \emph{International Conference on Social
	Computing, Behavioral-Cultural Modeling, and Prediction}.\hskip 1em plus
0.5em minus 0.4em\relax Springer, 2018.

\bibitem{alvari2016identifying}
H.~Alvari, A.~Hajibagheri, G.~Sukthankar, and K.~Lakkaraju, ``Identifying
community structures in dynamic networks,'' \emph{Social Network Analysis and
	Mining}, vol.~6, no.~1, p.~77, 2016.

\bibitem{ferrara2016predicting}
E.~Ferrara, W.-Q. Wang, O.~Varol, A.~Flammini, and A.~Galstyan, ``Predicting
online extremism, content adopters, and interaction reciprocity,'' in
\emph{International conference on social informatics}.\hskip 1em plus 0.5em
minus 0.4em\relax Springer, 2016, pp. 22--39.

\bibitem{alvari2011detecting}
H.~Alvari, S.~Hashemi, and A.~Hamzeh, ``Detecting overlapping communities in
social networks by game theory and structural equivalence concept,'' in
\emph{International Conference on Artificial Intelligence and Computational
	Intelligence}.\hskip 1em plus 0.5em minus 0.4em\relax Springer, 2011, pp.
620--630.

\bibitem{xiao2017wasserstein}
S.~Xiao, M.~Farajtabar, X.~Ye, J.~Yan, L.~Song, and H.~Zha, ``Wasserstein
learning of deep generative point process models,'' in \emph{Advances in
	Neural Information Processing Systems}, 2017, pp. 3247--3257.

\bibitem{tran2015netcodec}
L.~Tran, M.~Farajtabar, L.~Song, and H.~Zha, ``Netcodec: Community detection
from individual activities,'' in \emph{Proceedings of the 2015 SIAM
	International Conference on Data Mining}.\hskip 1em plus 0.5em minus
0.4em\relax SIAM, 2015, pp. 91--99.

\bibitem{xu2016learning}
H.~Xu, M.~Farajtabar, and H.~Zha, ``Learning granger causality for hawkes
processes,'' in \emph{International Conference on Machine Learning}, 2016,
pp. 1717--1726.

\bibitem{7490315}
V.~S. Subrahmanian, A.~Azaria, S.~Durst, V.~Kagan, A.~Galstyan, K.~Lerman,
L.~Zhu, E.~Ferrara, A.~Flammini, and F.~Menczer, ``The darpa twitter bot
challenge,'' 2016.

\bibitem{Cao:2014:ULG:2660267.2660269}
Q.~Cao, X.~Yang, J.~Yu, and C.~Palow, ``Uncovering large groups of active
malicious accounts in online social networks,'' in \emph{CCS}, 2014.

\bibitem{varol2017online}
O.~Varol, E.~Ferrara, C.~A. Davis, F.~Menczer, and A.~Flammini, ``Online
human-bot interactions: Detection, estimation, and characterization,''
\emph{ICWSM}, 2017.

\bibitem{ICWSM1715678}
T.~Green and F.~Spezzano, ``Spam users identification in wikipedia via editing
behavior,'' \emph{ICWSM}, 2017.

\bibitem{bessi2016}
A.~Bessi and E.~Ferrara, ``Social bots distort the 2016 us presidential
election online discussion,'' \emph{First Monday}, vol.~21, no.~11, 2016.

\bibitem{shao2017spread}
C.~Shao, G.~L. Ciampaglia, O.~Varol, A.~Flammini, and F.~Menczer, ``The spread
of fake news by social bots,'' \emph{arXiv preprint arXiv:1707.07592}, 2017.

\bibitem{ferrara2017}
E.~Ferrara, ``Disinformation and social bot operations in the run up to the
2017 french presidential election,'' 2017.

\bibitem{Zhu:2016:ISD:2942477.2942508}
K.~Zhu and L.~Ying, ``Information source detection in the sir model: A
sample-path-based approach,'' \emph{IEEE/ACM Trans. Netw.}, vol.~24, no.~1,
2016.

\bibitem{journals/corr/PeiMAZM14}
S.~Pei, L.~Muchnik, J.~S.~A. Jr., Z.~Zheng, and H.~A. Makse, ``Searching for
superspreaders of information in real-world social media.'' \emph{CoRR},
2014.

\bibitem{Fu2015}
H.~C.-Y. Fu, Yu-Hsiang and C.-T. Sun, ``Identifying super-spreader nodes in
complex networks,'' \emph{Mathematical Problems in Engineering}, 2015.

\bibitem{Cui:2013:COP:2487575.2487639}
P.~Cui, S.~Jin, L.~Yu, F.~Wang, W.~Zhu, and S.~Yang, ``Cascading outbreak
prediction in networks: A data-driven approach,'' in \emph{KDD}, 2013.

\bibitem{Konishi:2016:IKO:3061053.3061145}
T.~Konishi, T.~Iwata, K.~Hayashi, and K.-I. Kawarabayashi, ``Identifying key
observers to find popular information in advance,'' in \emph{IJCAI}, 2016.

\bibitem{benigni2016tweets}
M.~Benigni and K.~M. Carley, ``From tweets to intelligence: Understanding the
islamic jihad supporting community on twitter,'' in \emph{International
	Conference on Social Computing, Behavioral-Cultural Modeling and Prediction
	and Behavior Representation in Modeling and Simulation}.\hskip 1em plus 0.5em
minus 0.4em\relax Springer, 2016, pp. 346--355.

\bibitem{scanlonautomatic}
J.~R. Scanlon and M.~S. Gerber, ``Automatic detection of cyber-recruitment by
violent extremists,'' \emph{Security Informatics}, vol.~3, no.~1, p.~5, 2014.

\bibitem{scanlonforecasting}
------, ``Forecasting violent extremist cyber recruitment,'' \emph{IEEE
	Transactions on Information Forensics and Security}, vol.~10, no.~11, pp.
2461--2470, 2015.

\bibitem{hung2016detecting}
B.~W. Hung, A.~P. Jayasumana, and V.~W. Bandara, ``Detecting radicalization
trajectories using graph pattern matching algorithms,'' in \emph{Intelligence
	and Security Informatics (ISI), 2016 IEEE Conference on}.\hskip 1em plus
0.5em minus 0.4em\relax IEEE, 2016, pp. 313--315.

\end{thebibliography}

\end{document}